\documentclass[10pt,a4paper]{article}
\usepackage{amsmath,amssymb}
\usepackage{feynmf}
\usepackage[dvips]{epsfig}
\newcommand{\sgn}{\operatorname{sgn}}
\newcommand{\sech}{\operatorname{sech}}
\newcommand{\artanh}{\operatorname{artanh}}
\newcommand{\Li}{\operatorname{Li_2}}
\newcommand{\Tr}{\operatorname{Tr}}
\title{%
{\vspace{-3cm} \normalsize
     \hfill\parbox{38mm}{MS-TP-06-10 \\
                         cond-mat/0610376}  }\\[20mm]
The interfacial profile in two-loop order}

\author{Jens K\"uster and Gernot M\"unster%
\thanks{Institut f\"ur Theoretische Physik,
        Universit\"at M\"unster,
        Wilhelm-Klemm-Str.~9, D-48149 M\"unster, Germany;
        e-mail: munsteg@uni-muenster.de}}
\date{October 13, 2006\\(revised: July 27, 2007)}
\begin{document}
\maketitle

\begin{abstract}
The profile of interfaces separating different phases of statistical
systems is investigated in the framework of renormalized field theory. 
The profile function is calculated analytically in the local potential
approximation, using the effective potential to two loops. It can be
interpreted as an intrinsic interfacial profile. The loop corrections
to the leading tanh-type term turn out to be small.  They yield a
broadening of the interface.\\[5mm]
\end{abstract}
\section{Introduction}

Many systems of statistical physics can develop interfaces, separating
different phases or substances. These include Bloch walls in
ferromagnets, binary liquid mixtures,
liquid-gas systems and systems of immiscible polymers
\cite{Wi72,RW82,Bi83,Ja84,Safran}. The properties of interfaces are
important for the behaviour of such systems. They have been
investigated both experimentally and theoretically and are subject of
current research, see e.g.~\cite{Bi03}. For example, if substrates are covered by liquid
films in coatings or lubricants, the stability of the films depends on
the interfacial tensions between substrate, liquid and vapor. Another
example concerns the mechanical strength of polymer blends against
deformations. The structure of interfaces between domains of the
minority component, which are embedded in the background of the
majority component, is of crucial importance for the strength of the
blend \cite{Bi01}. 

The profile of an interface characterizes its geometry. It determines
several other properties such as the interface width and interface
tension and therefore plays a central role for the physics of
interfaces. Interfacial profiles are investigated experimentally
through scattering of photons or neutrons \cite{exp}. Theoretical
studies of interfacial profiles include numerical Monte Carlo
simulations \cite{St97,WSMB9799,MC,MM05} and field theoretic
calculations \cite{Ja84}.

The simplest description of interfaces is based on mean field theory
\cite{vdW93}. The interfacial profile is described by a continuous
order parameter $\phi(x)$ representing the difference between the
concentrations of the two coexisting phases. In the Landau-Ginzburg
approach the free energy density is written as \cite{Bellac}
\begin{equation}
\label{Lagr}
\mathcal{L}[\phi] = \frac{1}{2} \partial_\mu  \phi \partial_\mu  \phi +
V_0(\phi)\,.
\end{equation}
In the situation with interfaces the potential is of the double-well
type,
\begin{equation}
V_0(\phi) = \frac{g_0}{4!}\left(\phi^2 -v_0^2\right)^2.
\end{equation}
The minima of this potential correspond to the two homogeneous phases.
The mean field correlation length $\xi_0$ is defined through the second
moment of the correlation function in the mean field approximation.
It is given by the second derivative of the potential in its minima:
\begin{equation}
\xi_0^2 = \left( V''_0(v_0) \right)^{-1} = \frac{3}{g_0 v_0^2}\,.
\end{equation}

In mean field theory the interfacial profile is given by minimization
of the free energy density (\ref{Lagr}) with boundary conditions
appropriate for an interface. This leads to the differential equation
\begin{equation}
\Delta \phi - V'_0(\phi) = 0\,.
\end{equation}
If we choose the interface to be perpendicular to the $z$-axis, we find
the typical hyperbolic tangent profile \cite{CH58}
\begin{equation}
\phi(z) = v_0 \tanh \left(\frac{z-z_0}{2 \xi_0}\right).
\end{equation}
Its width is proportional to the mean field correlation length $\xi_0$.
The parameter $z_0$ specifies the location of the interface and will be
set to $0$ in the following.

Corrections to mean field theory come from fluctuations of the order
parameter field. They can be calculated systematically in renormalized
perturbation theory. The fluctuations result in different modifications
of the mean field result. First of all, higher order corrections change
the form of the profile from the tanh-function to a different function
$f$. Secondly, renormalization of the parameters $v_0$ and $\xi_0$
becomes necessary and consequently the mean field correlation length
$\xi_0$ is replaced by the physical correlation length $\xi$, which
diverges near the critical point with a characteristic exponent $\nu$.
A profile of the form
\begin{equation}
\label{form}
\phi(z) = v \, f ( z/2 \xi )
\end{equation}
corresponds to the scaling form proposed by Fisk and Widom \cite{FW60}.
Finally, long-wavelength capillary wave fluctuations lead to the
roughening phenomenon \cite{BLS65}. It implies a further broadening of
the interface such that in a three-dimensional system the width of an
interface depends logarithmically on the system size and diverges in
the limit of an infinite system.

In this article we consider corrections to mean field theory by means
of the ``local potential approximation", which is popular in the study
of non-homogeneous structures in field theory and cosmology. It takes
into account corrections to the local potential $V(\phi)$, whereas
contributions involving higher powers of derivatives are neglected. 
In this sense, the local potential approximation represents the leading
contributions in a systematic expansion of the effective action.
However, it neglects any non-local contribution to the effective
action. Therefore it is, in particular, insensitive to effects related
to the system size. The
resulting interfacial profile is of the scaling form (\ref{form}). In
the local potential approximation the effects of capillary waves are
not fully taken into account. This is due to the fact that in the
differential equation determining the interface profile only the local
value of the profile function but no large-scale properties like the
system size enter. The profile function therefore describes a kind of
intrinsic interfacial profile. Including large-scale effects would
require to build on the full effective potential, which is a non-local
functional of the profile function. In such a calculation additional
contributions to the interface profile are to be expected, which depend
logarithmically on the system size \cite{BLS65}. To define a unique
separation of an intrinsic profile from its capillary wave
contribution in such a situation is a nontrivial subject, which has
been discussed in various papers, see e.g.\
\cite{Wi72,We77,Ja84,WSMB9799,MM05}. In our approach, where broadening
due to capillary waves is suppressed, this problem does not show up.

Our calculations are performed in $D=3$ physical dimensions in contrast
to the $\epsilon$-expansions, where $D=4-\epsilon$ and an extrapolation
to $\epsilon=1$ is necessary. The three-dimensional approach, in the
framework of renormalized massive field theory, is based on a
systematic expansion in a dimensionless coupling \cite{Parisi80,LZ80}.
We calculate the potential to second order in the loop expansion.
Ultraviolet divergencies are treated by dimensional regularization
($D=3-\epsilon$), which does not vitiate the fact that the results for
physical quantities strictly refer to $D=3$ dimensions. Renormalization
of the three-dimensional field theory is performed in the scheme used
in \cite{Mu90} to two-loop order, employing the results of
\cite{MH94,GKM96}. The differential equation for the interfacial
profile is solved analytically in terms of special functions
\cite{Ku01}. The resulting profile is compared with results from Monte
Carlo calculations.

Previous studies of interfacial profiles in the framework of field
theory have been made in \cite{OK77,RJ78,JR78,SL86}. In
\cite{OK77,RJ78,SL86} the profile is calculated to first order in the
$\epsilon$-expansion. This amounts to a one-loop calculation. As the
expansion is around the four-dimensional case, capillary wave effects
are not included, and an extrapolation to $D=3$ is necessary. The
calculation of \cite{OK77} is extended to include the effects of an
external field in \cite{SL86}. In \cite{JR78} the interfacial profile
is considered in $D=3$ dimensions at one-loop order in the presence of
an external gravitational field. A functional form of the profile is
given, including capillary wave effects. The dependence on the system
size is, however, not considered.

\section{Effective potential}

The effective potential can be calculated in the loop-expansion by
standard field theoretic methods \cite{Ja74,IIM75,Amit}. The starting
point is the Hamiltonian density
\begin{equation}
h = \frac{1}{2} \partial_\mu  \phi \partial_\mu  \phi + V_0(\phi)\,.
\end{equation}
We write the mean field potential as
\begin{equation}
\label{mfpot}
V_0(\phi) = -\frac{m_{0}^2}4 \phi^2 + \frac{g_0}{4!}\phi^4
+\frac{3}{8} \frac{m_{0}^4}{g_0} 
= \frac{g_0}{4!}\left(\phi^2 -v_0^2\right)^2,
\end{equation}
where the bare mass $m_0 = 1 / \xi_0$ is equal to the inverse bare
(mean field) correlation length. The Hamiltonian is
\begin{equation}
\mathcal{H} =
\int\!\! d^D\!x\: h\,.
\end{equation}
In the physical situation the number of dimensions is of course $D=3$,
but we shall keep $D$ variable to allow for dimensional regularization.

The normalized partition function with an external source is defined
through the functional integral
\begin{equation}
Z[j] =  \frac{1}{Z_0} \int\!\! \mathcal{D} \phi(x)
\exp\left\{-\mathcal{H}[\phi] + \int\!\! d^D\!x \,j(x)\phi(x)\right\},
\end{equation}
where
\begin{equation}
Z_0 = \int\!\! \mathcal{D} \phi(x)
\exp\left\{-\mathcal{H}[\phi]\right\}\,,
\end{equation}
and the usual prefactor $\beta = 1 / kT$ is set to 1 by a suitable
normalization. The free energy is
\begin{equation}
W[j] = \ln Z[j]\,,
\end{equation}
and finally the Gibbs potential, which is called effective action in
field theory, is obtained by a Legendre transformation
\begin{equation}
\Gamma[\phi_c] = W[j] - \int\!\! d^D\!x\: j(x) \phi_c(x)\,,
\end{equation}
where the so-called ``classical field'' is
\begin{equation}
\phi_c(x) \doteq \frac{\delta W[j]}{\delta j(x)}\,.
\end{equation}

For a constant function $\phi_c(x) \equiv \phi$ the effective action
yields the effective potential $V(\phi)$ via
\begin{equation}
-\Gamma[\phi] = \int\!\! d^D\!x \, V(\phi)\,.
\end{equation}
More generally, in a derivative expansion the effective action can be
represented as
\begin{equation}
-\Gamma[\phi] = \int\!\! d^D\!x \,\left\{V(\phi(x))
+ \frac{1}{2} \mathcal{Z}(\phi(x))\,(\partial\phi(x))^2+\cdots\right\},
\end{equation}
where the dots imply terms with a higher number of derivatives, and
$\mathcal{Z}(\phi(x))$ is a field renormalization factor.

The effective potential $V(\phi)$ is the basic quantity for the
discussion of spontaneous symmetry breaking and non-homogeneous
configurations \cite{CW73}. It should be noted that it is not a convex
function, as the loop expansion also reveals. This is, however, not a
failure of the loop expansion, but a feature of the effective
potential, see e.g.\ \cite{WW87,Sher89} or sec.~2.2 of the textbook
\cite{MR07}. The local potential represents the density of the Gibbs
potential for a small volume element, in which the field is constant,
i.e.\ locally in a pure phase. The proof of convexity applies to the
extensive thermodynamic function, i.e.\ the total Gibbs potential, and
not to the local potential. For a proof of convexity in the context of
field theory see \cite{Symanzik70,IIM75}. In a situation, where the
local potential is not convex, but the Gibbs potential is, the system
is in a mixed phase and the convexity of the total Gibbs potential is
the result of the composition of contributions from the coexisting pure
phases. This is the situation where interfaces are present. Therefore
it is the local effective potential which is relevant for the local
interface structure. Another non-homogeneous situation, where the
non-convex local potential plays a role, is the presence of nucleation
bubbles in nucleation theory \cite{Langer69,MR00}.

The effective action can be calculated by
means of the loop expansion in the form
\begin{equation}
\Gamma[\phi] = 
\Gamma_0[\phi] + \Gamma_1[\phi] + \Gamma_2[\phi] + \cdots\,,
\end{equation}
where the leading term $\Gamma_0 = -\mathcal{H}$ is equal to the negative 
Hamiltonian. In the corresponding expansion of the effective potential,
\begin{equation}
V(\phi) = V_0(\phi) + V_1(\phi) + V_2(\phi) + \cdots\,,
\end{equation}
the leading term $V_0$ is the mean field potential (\ref{mfpot}).

The first order contribution to the effective action is
\begin{equation}
\Gamma_1[\phi] = -\frac{1}{2} \Tr \ln \frac{K(\phi)}{K(0)}
\end{equation}
with the differential operator
\begin{equation}
K(\phi) = - \Delta - \frac{m_{0}^2}{2} +\frac{g_0}{2}\phi^2\,.
\end{equation}
Setting $\phi$ constant, the first order correction to the effective
potential can be obtained by Fourier transform as
\begin{align}
V_{1}(\phi) & = \frac{1}{2} \int\!\!\frac{d^D k}{(2\pi)^D}\,
\ln\left( 1+ \frac{g_0 \phi^2/2}{k^2 -m_{0}^2/2}\right)\nonumber\\
& = \frac{1}{12\pi} \left\{m_{0}^3 
-\left(-\frac{m_{0}^2}{2} 
+\frac{g_0}{2}\phi^2\right)^\frac{3}{2}\right\}.
\end{align}

For small values of the field, $|\phi| < \sqrt{m_{0}^2 / g_0}$, this
contribution has an imaginary part. Actually, for small values of the
field the system is unstable due to spinodal decomposition.  The
boundary value for $|\phi|$ given above represents the mean field
spinodal. In the full renormalized theory this will be modified or even
be replaced by a crossover line \cite{Bi91}. According to \cite{WW87}
the imaginary part of the effective potential is to be identified with
the decay rate of the mixed state. For the calculation of the
interfacial profile the real part of the effective potential has to be
employed.

The two-loop contribution to the effective potential is obtained through
the calculation of two Feynman diagrams:
\begin{fmffile}{Graphs}
\begin{equation}
V_2(\phi) = -\frac{1}{8}\;
\raisebox{-9\unitlength}{\begin{fmfgraph}(24,20)
 \fmfpen{thick}
 \fmfsurroundn{o}{2}
 \fmf{plain,left=1}{o1,v1,o2,v1,o1}
 \fmfdot{v1}
\end{fmfgraph}}
-\frac{1}{12}\;
\raisebox{-9\unitlength}{\begin{fmfgraph}(32,20)
 \fmfsurroundn{o}{2}
 \fmf{dashes,tension=6,width=thin}{o1,v1}
 \fmf{dashes,tension=6,width=thin}{o2,v2}
 \fmfpen{thick}
 \fmf{plain,left=1}{v1,v2,v1}
 \fmf{plain}{v1,v2}
 \fmfdot{v1,v2}
\end{fmfgraph}}
\end{equation}
\end{fmffile}

In the two-loop contribution usual ultraviolet divergencies appear. They
originate from large momentum, short-distance fluctuations, and are not
related to capillary wave fluctuations. This is reflected in the fact
that they are identical to those coming from bulk fluctuations in a
situation without interfaces.

The divergencies have to be treated in some regularization
and renormalization scheme. We choose to employ dimensional
regularization in $D=3-\epsilon$ dimensions. It should be noted that
this does not amount to an $\epsilon$-expansion, since after
renormalization $\epsilon$ is sent to zero, whereas in the
$\epsilon$-expansion one has $D=4-\epsilon$ and the results have to be
extrapolated to $\epsilon = 1$. We obtain
\begin{align}
\label{V2}
V_{2}(\phi)&=
- \frac{g_{0}}{8\pi} \frac{m_{0}^2}{32 \pi}\nonumber\\
&\hspace*{-20pt} + \left(\frac{g_0}{8\pi}\right)^2 m_{0}^{-2\epsilon}\,
\frac{\phi^2}{6}\left\{- \frac{1}{\epsilon} 
+ \frac{1}{2} + \gamma + \ln\left(\frac{9}{4\pi}\right)
+\ln\left(-\frac{1}{2}+\frac{g_0\phi^2}{2m_{0}^2}\right)\right\},
\end{align}
where $\gamma=0.577\ldots$ is Euler's constant. This potential is being
used for the calculation of the interfacial profile.

\section{The Profile}

In the local potential approximation the effective action is
approximated by
\begin{equation}
-\Gamma_{\mathrm{LPA}}[\phi] = \int\!\! d^D\!x \,\left\{V(\phi(x))
+ \frac{1}{2 Z_3} (\partial\phi(x))^2 \right\}.
\end{equation}
The factor $Z_3$ is the usual field renormalization constant
in the broken symmetry phase, given by
\begin{equation}
Z_3 = \frac{1}{\mathcal{Z}(v)}\,,
\end{equation}
with $v$ denoting the minimum of the effective potential.

The interfacial profile $\phi(z)$ is now obtained in the local potential
approximation as a solution of the differential equation
\begin{equation}
Z_3^{-1} \Delta \phi - V_{r}'(\phi) = 0\,,
\end{equation}
where $V_{r}$ denotes the real part of the effective potential. For an
interface perpendicular to the $z$-axis the appropriate boundary
conditions are
\begin{align}
\lim_{z \to \infty} \phi(z) &= v \\
\lim_{z \to -\infty} \phi(z) &= -v.
\end{align}

The resulting profile function contains diverging coefficients coming
from the $1/\epsilon$ term in the second order effective potential
(\ref{V2}). These are treated by the usual renormalization procedure.
We adopt the renormalization scheme used in \cite{Mu90} to two-loop
order. The renormalized mass $m_R = 1 / \xi$ is equal to the inverse
correlation length $\xi$, which in turn is defined through the second
moment of the correlation function. The field $\phi$ and its value $v$
in the minimum are renormalized according to
\begin{equation}
\phi_R(x) = \frac{1}{\sqrt{Z_3}}\,\phi(x)\,, \qquad
v_R = \frac{1}{\sqrt{Z_3}}\,v\,.
\end{equation}
The renormalized coupling is specified as in \cite{LW87} through
\begin{equation}
g_R = \frac{3 m_R^2}{v_R^2}\,.
\end{equation}
In addition we define a dimensionless renormalized coupling according to
\begin{equation}
u_R = \frac{g_R}{m_R^{4-D}}\,.
\end{equation}

Employing the relations given in \cite{MH94,GKM96}, the bare quantities
$m_0$ and $g_0$ are expressed in terms of their renormalized
counterparts. This yields an expression for the renormalized
interfacial profile $\phi_R(z)$ depending on the parameters $m_R$ and
$u_R$. In this expression the divergencies are canceled, as they
should.

The solution is expanded in the form
\begin{align}
\phi_R(z) &=  \sqrt{\frac{3 m_R}{u_R}}
\left\{\chi_{0}(\hat{z}) + \frac{u_R}{8\pi}\chi_1(\hat{z}) 
+ \left(\frac{u_R}{8\pi}\right)^2 \chi_2(\hat{z})
+\mathcal{O}\left(u_R^3\right)\right\} \nonumber\\
&\equiv \sqrt{\frac{3 m_R}{u_R}} \, \chi(\hat{z})\,.
\end{align}
Here we have introduced the variable
\begin{equation}
\hat{z}=\frac{m_R}{2} z\,.
\end{equation}

The leading order contribution has of course the form of the
mean field profile
\begin{equation}
\chi_{0}(\hat{z}) = \tanh \hat{z}\,.
\end{equation}
It should, however, be distinguished from the mean field profile, as it
contains the physical correlation length instead of the mean field one.
It represents a refinement of Landau theory that is consistent with 
scaling \cite{FW60} and holds near the critical point.

Due to the fact that the effective potential has a point of non-analyticity,
the first and second order contributions are also non-analytic.
For $|\hat{z}|$ smaller or larger than $\artanh (1/\sqrt{3})$,
the functions $\chi_1$ and $\chi_2$ take different forms, which are called
$\chi_{1s}$, $\chi_{2s}$ and $\chi_{1l}$, $\chi_{2l}$, respectively.
For the first and second order correction we obtain \cite{Ku01}
\begin{equation}
\chi_{1s}(\hat{z})  
= \frac{1}{12}\hat{z} \sech^2 \hat{z} +\frac{2}{9}\sinh \hat{z} \cosh \hat{z}
-\frac{2}{3}\tanh \hat{z}
\end{equation}
\begin{align}
\chi_{1l}(\hat{z}) =
& \,\frac{1}{12}\hat{z} \sech^2 \hat{z}
-\frac{1}{12}\sech^2 \hat{z}\; \artanh\left(\frac{\sqrt{-\frac{1}{2}
+\sinh^2 \hat{z}}}{\sinh \hat{z}}\right) \nonumber \\
& +\frac{2}{9}\sinh \hat{z} \cosh \hat{z} -\frac{2}{3}\tanh \hat{z} \nonumber \\
& -\sinh \hat{z}\;\sqrt{-\frac{1}{2}+\sinh^2 \hat{z}}
\left(\frac{2}{9}-\frac{1}{2}\sech^2 \hat{z}\right),
\end{align}
and
{\allowdisplaybreaks
\begin{align}
\chi_{2s}(\hat{z}) =
& -\frac{1}{144}\hat{z}^2 \sech^2 \hat{z} \tanh \hat{z} \nonumber\\
& +\hat{z}\left(-\frac{13}{216} +\frac{17}{216}\cosh^2 \hat{z} 
-\frac{1}{24}\sinh^2 \hat{z} 
-\frac{815}{1728}\sech^2 \hat{z}\right) \nonumber\\
& +\frac{343}{648}\cosh \hat{z}\sinh \hat{z} +\frac{2}{81}\cosh^3 \hat{z}\sinh \hat{z}
-\frac{217}{216}\tanh \hat{z} \nonumber\\
& +\frac{1}{6}\sech^2 \hat{z} \left(\sqrt{3}-\artanh\frac{1}{\sqrt 3}\right)
\artanh\left(\sqrt{3}\tanh \hat{z}\right) \nonumber\\
& +\frac{1}{48}\sech^2 \hat{z} (\sinh(4\hat{z})-4\hat{z})
\ln\left(\frac{1}{2}-\frac{3}{2}\tanh^2 \hat{z}\right) \nonumber\\ 
& +\frac{1}{24}\sech^2 \hat{z} 
\left\{\Li\left(\frac{1-\sqrt{3}\tanh \hat{z}}{1-\sqrt{3}}\right)
-\Li\left(\frac{1-\sqrt{3}\tanh \hat{z}}{1+\sqrt{3}}\right)\right.\nonumber\\ 
& \left. \phantom{+\frac{1}{24}\sech^2 \hat{z}}
-\Li\left(\frac{1+\sqrt{3}\tanh \hat{z}}{1-\sqrt{3}}\right)
+\Li\left(\frac{1+\sqrt{3}\tanh \hat{z}}{1+\sqrt{3}}\right)\right\}
\end{align}
}
{\allowdisplaybreaks
\begin{align}
\chi_{2l}(\hat{z}) =
& \left(\frac{11\sqrt{3}}{108} -\frac{37}{108}\artanh\frac{1}{\sqrt 3}\right)
\sech^2 \hat{z} \sgn \hat{z} \nonumber\\
& -\frac{1}{144} \sech^2 \hat{z} \tanh \hat{z} \artanh^2 \left(\frac{\sqrt{-\frac{1}{2}
+\sinh^2 \hat{z}}}{\sinh \hat{z}}\right) \nonumber\\
& +\frac{1}{108}\artanh\left(\frac{\sqrt{-\frac{1}{2}
+\sinh^2 \hat{z}}}{\sinh \hat{z}}\right)
\left\{2 -4\cosh^2 \hat{z} +\frac{223}{16}\sech^2 \hat{z} \right.\nonumber\\
& \left. \hspace{10mm}
+\frac{3}{2}\hat{z} \sech^2 \hat{z} \tanh \hat{z} \right.\nonumber\\ 
& \left. \hspace{10mm}
+\left(4\cosh \hat{z}+\sech \hat{z}-9\sech^3 \hat{z}\right)\sqrt{-\frac{1}{2}
+\sinh^2 \hat{z}}\right\}\nonumber\\
& +\sqrt{-\frac{1}{2}+\sinh^2 \hat{z}} 
\left\{\frac{1}{108}\hat{z} 
\left(-4\cosh \hat{z}-\sech \hat{z} +9\sech^3 \hat{z}\right) \right.\nonumber\\ 
& \left. \hspace{10mm}
-\frac{233}{648}\sinh \hat{z}
-\frac{4}{81}\cosh^2 \hat{z} \sinh \hat{z} 
+\frac{145}{288}\sech \hat{z}\tanh \hat{z}\right\}\nonumber\\
& -\frac{1}{144}\hat{z}^2 \sech^2 \hat{z} \tanh \hat{z} \nonumber\\
& +\hat{z}\left(-\frac{13}{216} +\frac{17}{216}\cosh^2 \hat{z} 
-\frac{1}{24}\sinh^2 \hat{z}-\frac{223}{1728}\sech^2 \hat{z}\right) \nonumber\\
& +\frac{209}{648}\cosh \hat{z}\sinh \hat{z} +\frac{4}{81}\cosh^3 \hat{z}\sinh \hat{z}
\nonumber\\ 
& -\frac{19}{27}\tanh \hat{z} -\frac{3}{8}\sech^2 \hat{z} \tanh \hat{z} \nonumber\\
& +\frac{1}{6}\sech^2 \hat{z} \left(\sqrt{3}-\artanh\frac{1}{\sqrt 3}\right)
\artanh\left(\frac{1}{\sqrt{3}\tanh \hat{z}}\right) \nonumber\\
& +\frac{1}{48}\sech^2 \hat{z} \left(\sinh(4\hat{z})-4\hat{z}\right)
\ln\left(-\frac{1}{2}+\frac{3}{2}\tanh^2 \hat{z}\right) \nonumber\\ 
& +\frac{1}{24}\sech^2 \hat{z} 
\left\{\Li\left(\frac{1-\sqrt{3}\tanh \hat{z}}{1-\sqrt{3}}\right)
-\Li\left(\frac{1-\sqrt{3}\tanh \hat{z}}{1+\sqrt{3}}\right)\right.\nonumber\\
& \left. \phantom{+\frac{1}{24}\sech^2 \hat{z}}
-\Li\left(\frac{1+\sqrt{3}\tanh \hat{z}}{1-\sqrt{3}}\right)
+\Li\left(\frac{1+\sqrt{3}\tanh \hat{z}}{1+\sqrt{3}}\right)\right\}.
\end{align}
}

\begin{figure}[hbt]
\vspace{.8cm}
\centering
\epsfig{file=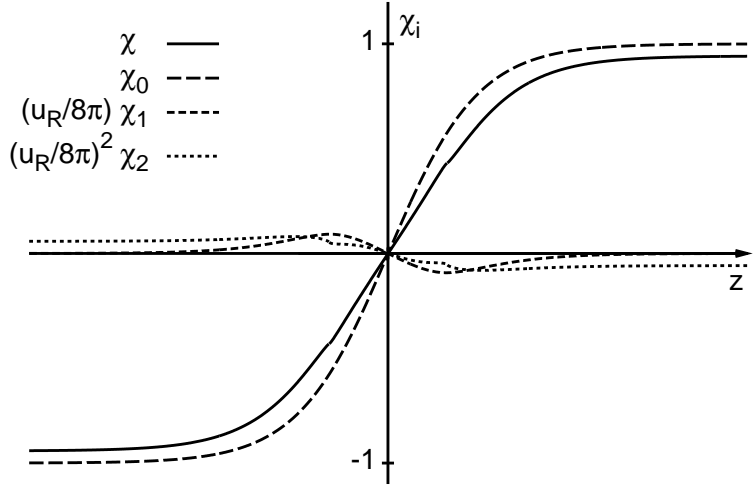,width=10cm}
\caption{\label{profile}
\small
The renormalized interfacial profile to second order.
The curves represent the contributions of the different orders
and their sum for the choice $u_R = 14.3$.}
\end{figure}

The contributions of the various orders and the resulting profile
function are displayed in Fig.\,\ref{profile}. For the renormalized
dimensionless coupling we have chosen $u_R = 14.3$. In the vicinity of
the critical point the coupling varies only slowly and is close to the
universal fixed point value $u_R^* = 14.3(1)$, see \cite{CH97} for a
discussion of numerical and field-theoretical estimates.

The figure shows that the corrections are small compared to the leading
term and therefore the interfacial profile at second order can be
approximately described by a tanh function. It can also be seen that
the higher order corrections lead to an increase of the interface
width. The broadening of the interface can be specified more
quantitatively in terms of a broadening factor $\alpha$ by fitting the
total profile function to
\begin{equation}
\chi_{\textrm{fit}}(\hat{z}) = 
\left(1 -\frac{1}{6} \left(\frac{u_R}{8\pi}\right)^2\right)\,
\tanh(\alpha\,\hat{z})\,,
\end{equation}
where the prefactor is fixed through the asymptotic behaviour 
of $\chi(\hat{z})$.
For $u_R = 14.3$ we obtain a value of $\alpha = 0.80$.

While the one-loop contribution $\chi_1$ vanishes asymptotically for
$|z| \rightarrow \infty$, the two-loop contribution approaches a small
but finite value, affecting the total asymptotic value. It also
displays a bump-like behaviour, which is due to the non-analyticity at
$\artanh (1/\sqrt{3})$, mentioned above.

A numerical investigation of the interfacial profile in the
three-di\-men\-sio\-nal Ising model has been made by Stauffer
\cite{St97} for a temperature, which is one percent below the critical
one. The resulting order parameter can be described rather well by a
tanh function and the numerical fit yields
\begin{equation}
\phi = 0.372\,\tanh(0.1\,z/a)\,,
\end{equation}
where $a$ denotes the lattice spacing.

In order to compare with our result we have to convert the
units by using
\begin{equation}
m_R = \frac{1}{\xi} 
= \frac{1}{a f_{-}} \left| \frac{T - T_c}{T_c} \right|^{\nu}
\end{equation}
with $f_-=0.2502(8)$ \cite{LF89} and $\nu=0.630$ \cite{GZ98,HPV99}. For
$T/T_c = 0.99$ this gives
\begin{equation}
0.80\,\frac{m_R}{2}z = 0.088\,\frac{z}{a} \,.
\end{equation}

The broadening factor deviates from the Monte Carlo value by only 12\%.
The significance of this coincidence is, however, not clear, due to the
procedure to identify the interface profile in the Monte Carlo
calculation. This procedure includes the effects of capillary waves,
but neglects the effects of bulk fluctuations, whose contributions are
central in our calculation. It is unclear how the resulting difference
in the definitions of the interface width affects the comparison.

\section{Conclusion}

We investigated the interfacial profile in the framework of
renormalized field theory using the local potential approximation. For
this purpose the effective potential has been calculated in the loop
expansion to second order. The resulting profile function is obtained
analytically. It is of the scaling form and can be interpreted as an
intrinsic profile. To lowest order the profile is of the mean field
type. For typical values of the coupling constant in the scaling region
the higher order corrections are small. They imply a broadening of the
interface of about 25\%.


\end{document}